\documentclass[a4paper,11pt]{amsart}
\usepackage[utf8]{inputenc}
\usepackage{amsaddr}
\usepackage{calrsfs}
\usepackage{graphicx,color}
\usepackage[english]{babel}
\usepackage{graphicx}
\usepackage{amsmath}
\usepackage{epstopdf}
\usepackage{float}
\usepackage{color}

\usepackage{amssymb}

\numberwithin{equation}{section}

\theoremstyle{definition}

\theoremstyle{plain}
\newtheorem{thm}{Theorem}[section]

\newtheorem{cor}{Corollary}[section]

\theoremstyle{definition}
\newtheorem{rem}{Remark}[section]

\begin{document}
\title[mixed fractional Merton model]
{The valuation of European option with transaction costs by mixed fractional Merton model}

\date{\today}

\author[Shokrollahi]{Foad Shokrollahi}
\address{Department of Mathematics and Statistics, University of Vaasa, P.O. Box 700, FIN-65101 Vaasa, FINLAND}
\email{foad.shokrollahi@uva.fi}

\begin{abstract}
This paper deals with the problem of discrete-time option pricing by the mixed
fractional version of Merton model with transaction costs. By a mean-self-financing delta hedging
argument in a discrete-time setting, a European call option pricing formula is obtained.
We also investigate the effect of the time-step $\delta t$ and the Hurst parameter $H$ on our pricing option model, which reveals that these parameters have high impact on option pricing. The properties of this model are also explained.
\end{abstract}
\keywords{Transaction costs;
mixed fractional Brownian motion;
European option;
Merton model}

\subjclass[2010]{91G20; 91G80; 60G22}

\maketitle

\section{Introduction}\label{sec:1}
Over the last few years, the financial markets have been regarded as complex and nonlinear dynamic systems. A series of studies has found that many financial market time series display scaling laws and long-range dependence. Therefore, it has been proposed that the Brownian motion in the classical Black-Scholes $(BS)$ model \cite{black} should be replaced by a process with long-range dependence.

Nowadays, the $BS$ model is the one most
commonly used for analyzing financial data, and some
scholars have presented modified forms of the $BS$ model which
have influential and significant outcomes on option pricing.
However, they are still theoretical adaptations and not necessarily
consistent with the empirical features of financial return
series, such as nonnormality, long-range dependence, etc.
For example, some scholars \cite{willinger,ozdemir,mariani,sottinen2003arbitrage,sottinen2016pathwise} have showed that returns are of long-range (or short-range) dependence, which suggests strong time-correlations
between different events at different time scales \cite{cajueiro,cajueiro2,mandelbrot}. In the search
for better models for describing long-range dependence in financial return series, a mixed fractional Brownian $(MFBM)$ model has been proposed as an improvement of the classical $BS$ model \cite{cheridito,mishura,mishura1,shokrollahi1,shokrollahi2,shokrollahi3,shokrollahi5,xiao,cont}. The advantage of
using the $MFBM$ is that the markets are free of arbitrage. Moreover, Cheridito \cite{cheridito}
has proved that, for $H\in(\frac{3}{4},1)$, the $MFBM$ is equivalent to one with Brownian motion, and hence time-step and long-range dependence in return series have no impact on
option pricing in a complete financial market without transaction costs. In addition, a number of empirical studies show that the paths of asset prices are discontinuous and that there are jumps in asset prices, both in the stock market and foreign exchange \cite{mandelbrot,merton,jarrow,ball,shokrollahi4}.

The above empirical findings have an important implication for option pricing.
Merton \cite{merton1} created a revolution in option pricing when the underlying asset was governed by a diffusion process. Based on this theory, Kou \cite{kou2002jump}, Cont and Tankov \cite{cont2002calibration} also considered the problems of pricing options under a jump diffusion environment in a larger setting. In this paper, to capture jumps or discontinuities, fluctuations
and to take into account the long memory property of financial markets, a mixed fractional version of the Merton model is introduced, which is based on a combination
of Poisson jumps and $MFBM$. The mixed fractional Merton $(MFM)$ model is based on the assumption that exchange rate returns are generated by a two-part stochastic process: (1)
small, continuous price movements are generated by a $MFBM,$ and (2) large, infrequent price
jumps are generated by a Poisson process. This two part
process is intuitively appealing, as it is consistent with
an impressive market in which major information arrives
infrequently and randomly. This process may
provide a description for empirically observed distributions
of exchange rate changes that are skewed, leptokurtic, have long
memory and fatter tails than comparable normal distributions
and apparent nonstationary variance.
Further, we will show the impact of the time-step and long-range dependence in return series exactly on option pricing, regardless of whether proportional transaction costs are considered or not in a discrete time setting.

Leland \cite{leland} is a pioneer scholar, who investigated option replication where transaction costs exist in a discrete time setting. In this view, the arbitrage-free arguments presented by Black and Scholes \cite{black} are not applicable in a model where transaction costs occur at all moments of trading of the stock or bond. The problem is that perfect replication incurs an infinite number of transaction costs because of the infinite variation which exists in the geometric Brownian motion. In this regard, a delta hedge strategy is constructed in accordance with revision conducted a discrete number of times. Transaction
costs lead to the failure of the no arbitrage principle and the continuous time trade in general: instead of no arbitrage, the principle of hedge pricing- according to which the price of an option is defined as the minimum level of initial wealth needed to hedge the option- comes into force .

According to the empirical findings obtained before and the views of behavioral finance and econophysics, we are motivated to examine the problem that exists in option pricing, while the dynamics of price  $S_t$ follows a mixed fractional jump-diffusion process under the transaction costs, we assume that $S_t$
satisfies

\begin{eqnarray}
S_t=S_0e^{\mu t+\sigma B(t)+\sigma_HB_H(t)+N_t\ln J}.
\label{eq:1}
\end{eqnarray}
where $S_0,\,\mu,\,\sigma$ and $\sigma_H$ are fixed, $B(t)$ is a Brownian motion, $B_H(t)$ is a fractional Brownian motion with Hurst parameter $H\in (\frac{3}{4},1)$, $N_t$ is a Poisson process with intensity $\lambda>0$, and $J$ is a positive random variable. Assume that $B(t),\,B_H(t),\,N_t$ and $J$ are independent.

This paper is organized into several sections. In Section \ref{sec:2}, we will study the problem of option pricing with transaction costs by applying delta hedging strategy. In addition, a new framework for pricing European option is obtained when the stock price $S_t$ is satisfied in equation  (\ref{eq:1}). Section \ref{sec:3} is devoted to empirical
studies and simulations to show the performance of the $MFM$ model. A conclusion is presented in Section \ref{sec:4}.\\

\section{Pricing option by mixed fractional version of Merton model with transaction costs}\label{sec:2}
Suppose $\{B(t)\}_{t\geq0}$ be a standard Brownian motion and $\{B_H(t)\}_{t\geq0}$ be a fractional Brownian motion with the Hurst parameter $H\in(\frac{3}{4},1)$, both defined on complete probability space $(\Omega,\mathcal{F},\mathcal{F}_t,,P)$, the absolute price jump size $J$ is a nonnegative random variable drawn from lognormal distribution, i.e. $\ln(J)=N(\mu_J,\sigma_J),$ which implies $$
J\sim Lognormal \Big(e^{\mu_J+\frac{\sigma_J^2}{2}},e^{2\mu_J+\sigma_J^2}(e^{\sigma_J^2}-1)\Big)$$ and a Poisson process $N=(N_t)_{t\geq0}$ with rate $\lambda$. Additionally, the processes $B, B_H, N$ and $J$ are independent, $P$ is the real world probability measure and  $(\mathcal{F}_t)_{t\in[0,T]}$ denotes the $P$-augmentation of filtration generated by $(B(\tau),B_H(\tau)),$ $\tau\leq t$.

The objective of this section is to derive a stock pricing formula under transaction costs in a discrete time setting. Consider $(D,S)$-market with a bond $D_t$ and a stock $S_t$, where
\begin{eqnarray}
D_t=D_0e^{rt}.
\label{eq:1-1}
\end{eqnarray}
and
\begin{eqnarray}
S_t=S_0e^{\mu t+\sigma B(t)+\sigma_HB_H(t)+N_t\ln J}\quad \mu, \sigma, \sigma_H\in R, D_0, S_0, t\in R^+ .
\label{eq:1-2}
\end{eqnarray}

The groundwork of modeling the effects of transaction costs was done by Leland \cite{leland}. He adopted the hedging strategy of
rehedging at every time-step $\delta t$. That is, with every $\delta t$ the portfolio is rebalanced, whether or not this is optimal in any sense. In the
following proportional transaction cost option pricing model, we follow the other usual assumptions in the Black-Scholes
model, but with the following exceptions:

\begin{enumerate}

\item[(i)] The price $S_t$ of the underlying stock at time $t$ satisfies equation (\ref{eq:1-2}).

\item[(ii)] The portfolio is revised every $\delta t$ where $\delta t$ is a finite and fixed, small time-step.

\item[(iii)]
Transaction costs are proportional to the value of the transaction in the underlying. Let $k$ denote the round trip
transaction cost per unit dollar of transaction. Suppose $U>0$ shares are bought $(U>0)$ or sold $(U<0)$ at the price $S_t$ , then
the transaction cost is given by
$\frac{k}{2}|U|S_t$ in either buying or selling, where $k$ is a constant. The value of $k$ will depend on the
individual investor. In the $MFM$ model, where transaction costs are incurred at every time the
stock or the bond is traded, the no arbitrage argument used by Black and Scholes no longer applies. The problem is that due
to the infinite variation of the $MFBM$, perfect replication incurs an infinite amount of transaction costs.

\item[(iv)] The hedge portfolio has an expected return equal to that from an option. This is exactly the same valuation policy as
earlier on discrete hedging with no transaction costs.

\item[(v)] Traditional economics assumes that traders are rational and maximize their utility. However, if their behavior is
assumed to be bounded rational, the traders' decisions can be explained both by their reaction to the past stock price, according
to a standard speculative behavior, and by imitation of other traders' past decisions, according to common evidence in
social psychology. It is well known that the delta hedging strategy plays a central role in the theory of option pricing and
that it is popularly used on the trading floor. Based on the availability heuristic, suggested by Tversky and Kahneman \cite{tversky},
traders are assumed to follow, anchor, and imitate the Black-Scholes delta hedging strategy to price an option.

\end{enumerate}

Let the price of European call option be denoted with expiration $T$ and strike price $K$ by $C(t,S_t)$ with boundary conditions:
\begin{eqnarray}
C(T,S_T)=(S_T-K)^+, \quad C(t,0)=0,\quad C(t,S_t)\rightarrow S_t\quad as \, S_t\rightarrow\infty.
\label{eq:2}
\end{eqnarray}
Then, $C(t,S_t)$ is derived by the following theorem.\\

\begin{thm} The price at every $t\in [0,T]$ of a European call option with strike price $K$ that matures at time $T$ is
given by
\begin{eqnarray}
C(t,S_t)=\sum_{n=0}^\infty\frac{e^{-\lambda'(T-t)}(\lambda'(T-t))^n}{n!}\Big[S_t\phi(d_1)-Ke^{-r(T-t)}\phi(d_2)\Big].
\label{eq:5}
\end{eqnarray}
Moreover, $C(t,S_t)$ satisfies the following equation
\begin{eqnarray}
&&\frac{\partial C}{\partial t}+rS_t\frac{\partial C}{\partial S_t}+\frac{S_t^2\widehat{\sigma}^2}{2}\frac{\partial^2 C}{\partial S_t^2}-rC\nonumber\\
&&+\lambda E[C(t,JS_t)-C(t,S_t)]-\lambda E[J-1]S_t\frac{\partial C}{\partial S_t}=0,
\label{eq:5-1}
\end{eqnarray}
where
\begin{eqnarray}
d_1=\frac{\ln\left(\frac{S_t}{K}\right)+r_n(T-t)+\frac{\sigma_n^2}{2}(T-t)}{\sigma_n\sqrt{T-t}},\quad d_2=d_1-\sigma_n\sqrt{T-t},
\label{eq:6}
\end{eqnarray}
\begin{eqnarray}
\lambda'=\lambda E(J)=\lambda e^{\mu_J+\frac{\sigma_J^2}{2}},\quad{\sigma_n}^2=\widehat{\sigma}^2+\frac{n\sigma_J^2}{T-t},
\label{eq:7}
\end{eqnarray}
\begin{eqnarray}
r_n&=&r-\lambda E(J-1)+\frac{n\ln E(J)}{T-t}\nonumber\\
&=&r-\lambda (e^{\mu_J+\frac{\sigma_J^2}{2}}-1)+\frac{n(\mu_J+\frac{\sigma_J^2}{2})}{T-t},
\label{eq:8}
\end{eqnarray}
\begin{eqnarray}
\widehat{\sigma}^2=\sigma^2+\sigma_H^2(\delta t)^{2H-1}+k\sqrt{\frac{2}{\pi}\Big(\frac{\sigma^2}{\delta t}+\sigma_H^2(\delta t)^{2H-2}\Big)}sign(\Gamma),
\label{eq:9}
\end{eqnarray}
sign$(\Gamma)$ is the signum function of $\frac{\partial^2C}{\partial S_t^2}$, $n$ is the number of prices jumps, $\delta t$ is a small and fixed time-step, $k$ is the transaction costs and $\phi(.)$ is the cumulative normal distribution.
\label{th:3-1}
\end{thm}

Moreover, using the put–call parity, we can easily obtain the valuation model for a put currency option, which is provided
by the following corollary.
\begin{cor} The value of European put option with transaction costs is given by

\begin{eqnarray}
P(t,S_t)=\sum_{n=0}^\infty\frac{e^{-\lambda'(T-t)}(\lambda'(T-t))^n}{n!}\Big[Ke^{-r(T-t)}\phi(-d_2)-S_t\phi(-d_1)\Big]\nonumber.
\label{eq:10}
\end{eqnarray}
\end{cor}

\section{Properties of pricing formula}\label{sec:3}

In this section, we present the properties of $MFM$ 's log-return density. The effects of Hurst parameter and time-step on our modified volatility $(\sigma^2_n)$ are also discussed  in the discrete time and continuous time cases. Then we show that these parameters play a significant role in a discrete time setting, both with and without transaction costs.
\subsection{Log-return density}

In the case of $MFM$ the log return jump size is assumed to be $(Y_i)=(\ln J_i)\sim N(\mu_J,\sigma_J^2)$ and the probability density of log return $x_t=\ln (S_t/S)$ is achieved as a quickly converging series of the following form:

\begin{eqnarray}
P(x_t\in A)&=&\sum_{n=0}^\infty P(N_t=n)P(x_n\in A)|N_t=n)\nonumber\\
P(x_t)&=&\sum_{n=0}^\infty\frac{e^{-\lambda t}(\lambda t)^n}{n!}N(x_t;\mu t+n\mu_J,\sigma^2t+\sigma_H^2t^{2H}+n\sigma_J^2),
\label{eq:10-1}
\end{eqnarray}
where $N(x_t;\mu t+n\mu_J,\sigma^2t+\sigma_H^2t^{2H}+n\sigma_J^2)$
\begin{eqnarray}
=\frac{1}{\sqrt{2\pi(\sigma^2t+\sigma_H^2t^{2H}+n\sigma_J^2)}}\exp\Big[-\frac{(x_t-(\mu t+n\mu_J)
)^2}{2(\sigma^2t+\sigma_H^2t^{2H}+n\sigma_J^2)}\Big]
\label{eq:10-2}
\end{eqnarray}

The term $P(N_t=n)=\frac{e^{-\lambda t}(\lambda t)^n}{n!}$ is the probability that the asset price jumps $n$ times during the time interval of length $t$. And $P(x_n\in A)|N_t=n)=N(x_t;\mu t+n\mu_J,\sigma^2t+\sigma_H^2t^{2H}+n\sigma_J^2)$
is the mixed fractional normal density of log-return. It supposes that the asset price jumps $i$ times in the time interval of $t$. As a result, in the $MFM$ model, the log-return density can be described as the weighted average of the mixed fractional normal density by the probability that the asset price jumps $n$ times.\\

\begin{figure}[H]
  \centering
          \includegraphics[width=1\textwidth]{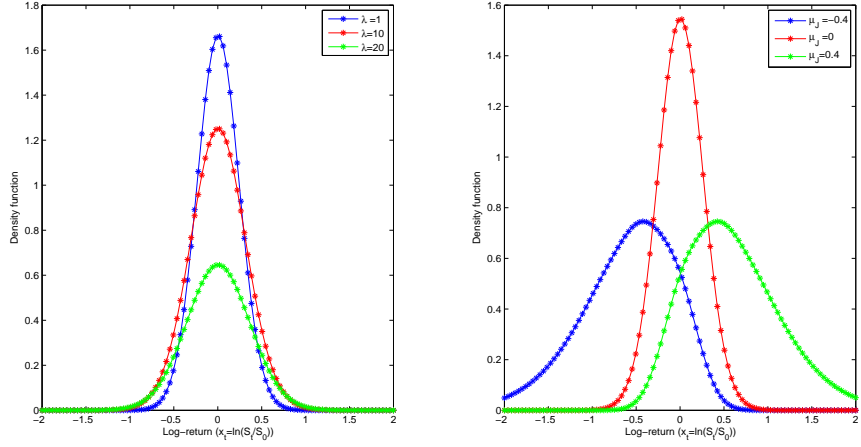}

  \caption{$MFM$'s Log-Return Density. Fixed parameters are $\sigma=0.25,\sigma_H=0.25,H=0.76,\sigma_J=0.1,\lambda=3,\mu_J=0,\mu=0.009$, and
$t=0.5.$}
\label{fig:1}
\end{figure}

 The outstanding properties of log-return density $P(x_t)$ are observed in the $MFM$. Firstly, the  $\mu_J$ sign refers to the expected log-return jump size, $E(Y)=E(\ln J)=\mu_J$, which indicates the skewness sign. If $\mu_J<0$, the log-return density $P(x_t)$ shows negatively skewed, and if $\mu_J=0$, it is  symmetric as displayed on the right side of Figure \ref{fig:1}.

Secondly, a larger value of intensity $\lambda$ (i.e., the expectation is toward the frequent occurrences of jumps) leads to fatter-tailed density, as indicated on the left side of Figure \ref{fig:1}. It is significant that the case  $\lambda=20$ is much smaller than when $\lambda=1$ or $\lambda=10$ due to the fact that  excess kurtosis is identified as a standardized measure (by standard deviation). Tables \ref{table:1} and \ref{table:2} show the annualized moments of $MFM$'s Log-Return density on the right side and left side of Figure \ref{fig:1}, respectively.
\begin{center}
\begin{table}[H] \caption{ Moments of $MFM$'s Log-Return density}

\begin{tabular}{|ccccc|} \hline
Model & Mean& Standard Deviation& Skewness& Excess Kurtosis \\
\hline
$\mu_J=-0.4$ &    -1.1910  &  0.6161 & -0.5082  & 0.2806\\
$\hspace{-6mm}\mu_J=0   $    &     0.0090  &  0.1361 &   0      &  0.706\\
$\hspace{-4mm}\mu_J=0.4$  &    1.2090   &  0.6161 & 0.5082  & 0.2806\\

\hline

\end{tabular}
\label{table:1}
\end{table}
\end{center}

\begin{center}
\begin{table}[H] \caption{ Moments of $MFM$'s Log-Return density }

\begin{tabular}{|ccccc|} \hline
Model & Mean& Standard Deviation& Skewness& Excess Kurtosis \\
\hline
$\lambda=1$     &    0.0040   &  0.1161 &   0   &  0.0223\\
$\lambda=10$    &    -0.0411  &  0.2061 &   0   &  0.706\\
$\lambda=20$    &    -0.0913  &  0.3061 &    0  &  0.0640\\

\hline

\end{tabular}
\label{table:2}
\end{table}
\end{center}

\subsection{The impact of parameters}

Mantegna and Stanley \cite{mantegna} as pioneer scholars proposed the scaling invariance method  from the complex science of economic
systems which led to numerous investigations into scaling laws in finance. The major question in economics is whether the price impact of scaling law and long-range dependence is significant in option pricing. The answer to this question is assured. For instance, one of the significant issues in finance concerning the modeling of high-frequency data is related to analyzing the volatility in different time scales.

\begin{rem} In a continuous time setting $(\delta t=0, \lambda\neq 0)$ without transaction costs the implied volatility is $\widehat{\sigma}_n^2=\sigma^2+\frac{n\sigma_J^2}{T-t}$, thus the option value is similar to the Merton jump diffusion model \cite{merton}. Moreover, if $\delta t=0$ in the absence of transaction costs and jump case, the $MFM$ model reduces to the $BS$ model
\begin{eqnarray}
\frac{\partial C}{\partial t}+rS_t\frac{\partial C}{\partial S_t}+\frac{S_t^2\sigma^2}{2}\frac{\partial^2 C}{\partial S_t^2}-rC=0,
\label{eq:11}
\end{eqnarray}
which shows that the Hurst parameter $H$  and time-step $\delta t$ have no effect on option pricing model in a continuous time setting $(\delta t=0)$.
\end{rem}

\begin{figure}[H]
  \centering
          \includegraphics[width=1\textwidth]{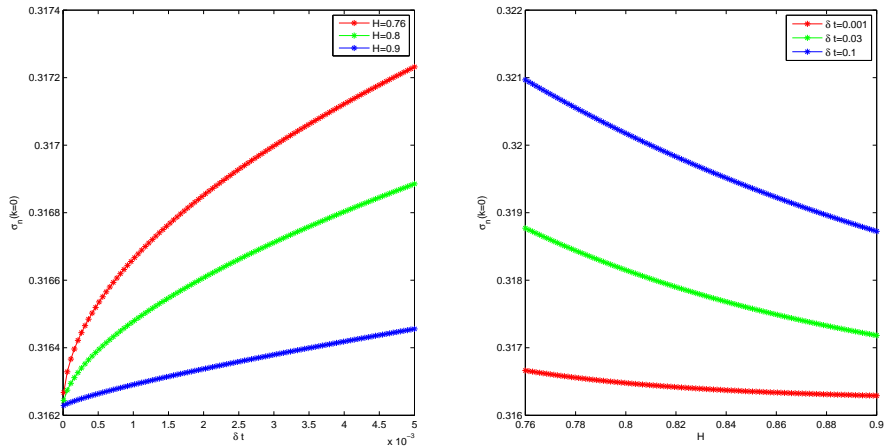}

  \caption{Modified volatility. Fixed parameters are $\sigma=0.1,\sigma_H=0.1,H=0.76,\sigma_J=0.03,T=0.2,k=0$, and
$t=0.1$}
\label{fig:2}
\end{figure}

\begin{figure}[H]
  \centering
          \includegraphics[width=1\textwidth]{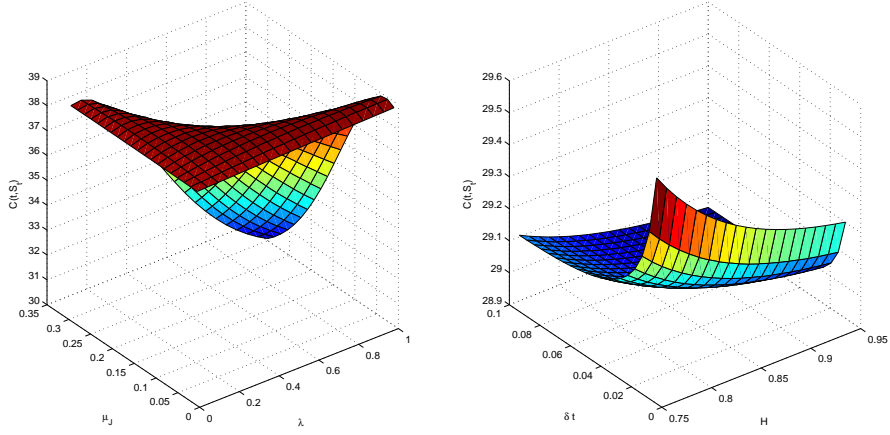}

  \caption{European Call option. Fixed parameters are $\sigma=0.1,\sigma_H=0.3,H=0.76,\sigma_J=0.01,T=2,k=0.1,K=135,S_t=140,\delta t=0.03$, and
$t=0.05$}
\label{fig:3}
\end{figure}

\begin{rem} In a discrete time setting without transaction costs $(k=0$ ,$\delta t\neq0)$, if jump occurs, the modified volatility is $\widehat{\sigma}_n^2=\sigma^2+\sigma_H^2(\delta t)^{2H-1}+\frac{n\sigma_J^2}{T-t}$  and when jump does not occur $(\lambda=0)$, from equation (\ref{eq:5-1}), we have

\begin{eqnarray}
\frac{\partial C}{\partial t}+rS_t\frac{\partial C}{\partial S_t}+(\sigma^2+\sigma_H^2 (\delta t)^{2H-1})\frac{S_t^2}{2}\frac{\partial^2 C}{\partial S_t^2}-rC=0,
\label{eq:11}
\end{eqnarray}
which demonstrates that the delta hedging strategy in a discrete time case is fundamentally different in comparison with a continuous time case. It also indicates that the scaling  exponent $2H-1$  and time-step $\delta t$ play a significant role in option pricing theory. Figure \ref{fig:2} illustrates the impacts of Hurst parameter and time-step on modified volatility. Moreover, the impacts of the time-step, Hurst parameter, mean jump, and jump intensity on our European call option are shown in Figure \ref{fig:3}.
\end{rem}

\begin{rem}  From \cite{wang1} we infer there exists $\delta t\in (0,\frac{1}{M})$ such that
\begin{eqnarray}
\min_{\delta t\in (0,\frac{1}{M})}\widehat{\sigma}^2,
\label{eq:12}
\end{eqnarray}
holds,

where $M>1$, $k$ is small enough

\begin{eqnarray}
\widehat{\sigma}^2=\sigma^2+\sigma_H^2(\delta t)^{2H-1}+k\sqrt{\frac{2}{\pi}\left(\frac{\sigma^2}{2\delta t}+\sigma_H^2(\delta t)^{2H-2}\right)}.
\label{eq:13}
\end{eqnarray}
Indeed,
\begin{eqnarray}
\sigma_H^2(\delta t)^{2H-1}+k\sqrt{\frac{2}{\pi}\left(\frac{\sigma^2}{\delta t}+\sigma_H^2(\delta t)^{2H-2}\right)}\nonumber\\
\geq 2\sigma_H(\delta t)^{H-\frac{1}{2}}k^{\frac{1}{2}}\left(\frac{2}{\pi}\left(\frac{\sigma^2}{\delta t}+\sigma_H^2(\delta t)^{2H-2}\right)\right)^{\frac{1}{4}}.
\label{eq:14}
\end{eqnarray}
Set
\begin{eqnarray}
\sigma_H^2(\delta t)^{2H-1}=\sqrt{\frac{2k^2}{\pi}\left(\frac{\sigma^2}{\delta t}+\frac{\sigma_H^2}{(\delta t)^{2-2H}}\right)}.
\label{eq:15}
\end{eqnarray}
Thus
\begin{eqnarray}
\sigma_H^2(\delta t)^{2H}=\frac{\frac{2k^2}{\pi}+\sqrt{\left(\frac{2k^2}{\pi}\right)^2+\frac{8k^2}{\pi}\sigma^2\delta t}}{2}.
\label{eq:16}
\end{eqnarray}
Suppose
\begin{eqnarray}
f(x)=\sigma_H^2x^{2H}-\frac{\frac{2k^2}{\pi}+\sqrt{\left(\frac{2k^2}{\pi}\right)^2+\frac{8k^2}{\pi}\sigma^2x}}{2}.
\label{eq:17}
\end{eqnarray}
Since $f(0)<0$ and
\begin{eqnarray}
f\left(\frac{1}{M}\right)=\sigma_H^2\left(\frac{1}{M}\right)^{2H}-\frac{\frac{2k^2}{\pi}+\sqrt{\left(\frac{2k^2}{\pi}\right)^2
+\frac{8k^2}{\pi}\sigma^2\frac{1}{M}}}{2}>0,
\label{eq:18}
\end{eqnarray}
as $k$ is small enough.

Hence, there exists a $\delta t\in (0,\frac{1}{M})$ such that $\min_{\delta t\in (0,\frac{1}{M})}\widehat{\sigma}^2$ holds.

Suppose
\begin{eqnarray}
\widehat{\sigma}^2(\min)=\min_{\delta t\in (0,\frac{1}{M})}\widehat{\sigma}^2,
\label{eq:19}
\end{eqnarray}
so
\begin{eqnarray}
\sigma_n^2(\min)=\min_{\delta t\in (0,\frac{1}{M})}\sigma_n^2=\min_{\delta t\in (0,\frac{1}{M})}\widehat{\sigma}^2+\frac{n\delta^2}{T-t}.
\label{eq:20}
\end{eqnarray}
Then the minimal price of an option with respect to transaction costs is displayed as $C_{\min}(t,S_t)$ with $\sigma_n^2(\min)$ in equation (\ref{eq:5}).
$C_{\min}(t,S_t)$ can be applied to the real price of an option.
\end{rem}

\section{Conclusion}\label{sec:4}

Without using the arbitrage argument, in order to capture jumps or discontinuities, fluctuations,
and to take into account the long memory property, this paper obtains the $MFM$ model by delta hedging
strategy in discrete time setting.
Some properties of $MFM$'s log-return density are discussed. Moreover, we infer that the Hurst parameter $H$ and time-step $\delta t$ play a significant role in pricing option in a discrete time setting for cases both with and without transaction costs.
But these parameters have no impact on option pricing in a continuous time
setting.

\section*{Appendix}\label{appendix}
\textbf{Proof of Theorem \ref{th:3-1}.} We consider a replicating portfolio with $\psi(t)$ units of financial underlying asset and one unit of the option. Then, the value of the portfolio at time $t$ is
\begin{eqnarray}
P_t=\psi(t)S_t-C(t,S_t).
\label{eq:a1}
\end{eqnarray}

Now, the movement in $S_t$ and $P_t$ is considered under discrete time interval $\delta t$.
 In view of this, we suppose that trading takes place at the specific time points of $t$ and $t+\delta t$. It can be said that the number of shares through the use of delta-hedging strategy and the present stock price $S_t$ are constantly held during the rebalancing interval $[t,t+\delta t)$. Then, the movement in the value of the portfolio after time interval $\delta t$ is defined as follows:

\begin{eqnarray}
\delta P_t=\psi(t)\delta S_t-\delta C(t,S_t)-\frac{k}{2}|\delta\psi(t)|S_t.
\label{eq:a2}
\end{eqnarray}

Where $\delta S_t$ is the movement of the underlying stock price, $\delta \psi(t)$ is the movement of the number of units of  stock held in the portfolio, and $\delta P_t$ is the change in the value of the portfolio.

Since the time-step $\delta t$ and the asset change are both small, according to Taylor's formulae we have if $\delta N_t=0$ with probability $1-\lambda\delta t$, so

\begin{eqnarray}
\delta S_t&&=S_t\mu\delta t+S_t\delta \sigma B(t)+S_t\delta\sigma_H B_H(t)+\frac{S_t}{2}\big(\mu\delta t +\sigma\delta B(t)+\sigma_H\delta B_H(t)\big)^2 \nonumber \\
&&+\frac{S_t}{6}e^{\theta[\mu\delta t+\sigma\delta B(t)+\sigma_H\delta B_H(t)]}\big(\mu\delta t+\sigma\delta B(t)+\sigma_H\delta B_H(t)\big)^3,
\label{eq:a3}
\end{eqnarray}
where $\theta=\theta(t,w),\quad w\in\Omega$, and $0<\theta<1$.

Since $B(t)$ and  $B_H(t)$ are continuous, then from \cite{wang} we have

\begin{eqnarray}
(\delta t)\delta B_H(t)=O\Big((\delta t)^{1+H}\sqrt{\log \frac{1}{\delta t}}\Big),
\label{eq:a4}
\end{eqnarray}

\begin{eqnarray}
(\delta t)\delta B(t)=O\Big((\delta t)^{\frac{3}{2}}\sqrt{\log \frac{1}{\delta t}}\Big),
\label{eq:a5}
\end{eqnarray}

\begin{eqnarray}
\frac{\delta B_H(t)}{\delta B(t)}\rightarrow 0 \qquad as\quad \delta t\rightarrow 0,
\label{eq:a6}
\end{eqnarray}
and
\begin{eqnarray}
&&e^{\theta[\mu\delta t+\sigma \delta B(t)+\sigma_H\delta B_H(t)]}[\mu\delta t+\sigma \delta B(t)+\sigma_H\delta B_H(t)]^3\nonumber \\
&&=O((\delta t)^3)+O\big((\delta t)^{\frac{5}{2}}\sqrt{\log (\delta t)^{-1}}\big)+O\big((\delta t)^2\log(\delta t)^{-1}\big)+O\big((\delta t)^{\frac{3}{2}}(\log (\delta t)^{-1})^{\frac{3}{2}}\big)\nonumber \\
&&=O\big((\delta t)^{\frac{3}{2}}(\log (\delta t)^{-1})^{\frac{3}{2}}\big)\nonumber.
\label{eq:a7}
\end{eqnarray}

Thus, we can get
\begin{eqnarray}
\delta S_t&=&\mu S_t\delta t+ S_t[\sigma \delta B(t)+\sigma_H\delta B_H(t)]\nonumber \\
&+&\frac{S_t}{2}[\sigma \delta B(t)+\sigma _H \delta B_H(t)]^2+O\big((\delta t)^{\frac{3}{2}}\sqrt{\log (\delta t)^{-1}}\big),
\label{eq:a8}
\end{eqnarray}
\begin{eqnarray}
(\delta S_t)^2=S_t^2[\sigma\delta B(t)+\sigma_H\delta B_H(t)]^2+O\big((\delta t)^{\frac{3}{2}}\sqrt{\log (\delta t)^{-1}}\big),
\label{eq:a9}
\end{eqnarray}

\begin{eqnarray}
\delta C(t,S_t)=\frac{\partial C}{\partial t}\delta t+\frac{\partial C}{\partial S_t}\delta (S_t)+\frac{1}{2}\frac{\partial^2C}{\partial S_t^2}(\delta S_t)^2+O\big((\delta t)^{\frac{3}{2}}\sqrt{\log (\delta t)^{-1}}\big),
\label{eq:a10}
\end{eqnarray}
and
\begin{eqnarray}
\delta \psi(t)=\frac{\partial \psi}{\partial t}\delta t+\frac{\partial \psi}{\partial S_t}\delta S_t+\frac{1}{2}\frac{\partial^2\psi}{\partial S_t^2}(\delta S_t)^2+O\big((\delta t)^{\frac{3}{2}}\sqrt{\log (\delta t)^{-1}}\big).
\label{eq:a11}
\end{eqnarray}
If $\delta N_t=1$ with probability $\lambda\delta t$ and the jump of $N_t$ in $[t,t+\Delta t]$ is assumed to occur at current time $t$, then

\begin{eqnarray}
S_{t^+}=S_0e^{\mu t+\sigma B(t)+\sigma_HB_H(t)+\ln J},
\label{eq:a12}
\end{eqnarray}
\begin{eqnarray}
S_{t+\delta t}=S_0e^{\mu (t+\delta t)+\sigma B(t+\delta t)+\sigma_HB_H(t+\delta t)+\ln J},
\label{eq:a13}
\end{eqnarray}
\begin{eqnarray}
\delta S_{t^+}=S_{t+\delta t}-S_{t^+}=S_{t^+}\big[e^{\mu t+\sigma B(t)+\sigma_HB_H(t)}-1\big],
\label{eq:a14}
\end{eqnarray}
\begin{eqnarray}
\delta S_t&=&S_{t+\delta t}-S_t=S_{t+\delta t}-S_{t^+}+S_{t^+}-S_t\nonumber\\
&=&S_{t^+}\big[e^{\mu t+\sigma B(t)+\sigma_HB_H(t)}-1\big]+(S_{t^+}-S_t)
\label{eq:a15}
\end{eqnarray}

\begin{eqnarray}
\delta C(t,S_t)&=&C(t+\delta t,S_{t+\delta t})-C(t,S_{t^+})+C(t,S_{t^+})-C(t,S_t)\nonumber\\
&=&C(t,S_{t^+})-C(t,S_t)+\frac{\partial C}{\partial t}\delta t+\frac{\partial C}{\partial S_{t^+}}\delta (S_{t^+})\nonumber\\
&+&\frac{1}{2}\frac{\partial^2C}{\partial S_{t^+}^2}(\delta S_{t^+})^2+O\big((\delta t)^{\frac{3}{2}}\sqrt{\log (\delta t)^{-1}}\big),
\label{eq:a16}
\end{eqnarray}

\begin{eqnarray}
\delta \psi(t,S_t)&=&\psi(t,S_{t^+})-\psi(t,S_t)+\frac{\partial \psi(t,S_{t^+})}{\partial t}\delta t+\frac{\partial \psi(t,S_{t^+})}{\partial S_{t^+}}\delta (S_{t^+})\nonumber\\
&+&\frac{1}{2}\frac{\partial^2\psi(t,S_{t^+})}{\partial S_{t^+}^2}(\delta S_{t^+})^2+O\big((\delta t)^{\frac{3}{2}}\sqrt{\log (\delta t)^{-1}}\big),
\label{eq:a17}
\end{eqnarray}
where $\delta S_{t^+}=S_{t+\delta t}-S_{t^+}$.

Based on the above assumptions iv and v, we have $E\Big(\frac{\delta P_t}{P_t}\Big)=\frac{\delta D_t}{D_t}$, i.e. $E\delta P_t=rP_t+O\big((\delta t)^2\big)$. Then
\begin{eqnarray}
&&(1-\lambda\delta t)E\Big[\psi\delta S_t-\delta C(t,S_t)-\frac{kS_t}{2}|\delta \psi(t)|\Big]\nonumber\\
&+&\lambda\delta tE\Big[S_{t^+}\big(e^{\mu t+\sigma B(t)+\sigma_HB_H(t)}-1\big)\psi(t)\nonumber\\
&+&(S_{t^+}-S_t)\psi(t)-\Big(C(t,S_{t^+})-C(t,S_t)+\frac{\partial C}{\partial t}\delta t+\frac{\partial C}{\partial S_{t^+}}\delta (S_{t^+})\nonumber\\
&+&\frac{1}{2}\frac{\partial^2C}{\partial S_{t^+}^2}(\delta S_{t^+})^2+O\big((\delta t)^{\frac{3}{2}}\sqrt{\log (\delta t)^{-1}}\big)\Big)-\frac{kS_t}{2}\Big|\psi(t,S_{t^+})-\psi(t,S_t)\nonumber\\
&+&\frac{\partial \psi(t,S_{t^+})}{\partial t}\delta t+\frac{\partial \psi(t,S_{t^+})}{\partial S_{t^+}}\delta (S_{t^+})
+\frac{1}{2}\frac{\partial^2\psi(t,S_{t^+})}{\partial S_{t^+}^2}(\delta S_{t^+})^2\nonumber\\
&+&O\big((\delta t)^{\frac{3}{2}}\sqrt{\log (\delta t)^{-1}}\big)\Big]=rP_t\delta t\nonumber,
\label{eq:a18}
\end{eqnarray}
i.e.
\begin{eqnarray}
&&(1-\lambda\delta t)E\Big[\psi\delta S_t-\delta C(t,S_t)\Big]+\lambda\delta tE\Big[S_{t^+}\big(e^{\mu t+\sigma B(t)+\sigma_HB_H(t)}-1\big)\psi(t)\nonumber\\
&+&(S_{t^+}-S_t)\psi(t)-\Big(C(t,S_{t^+})-C(t,S_t)+\frac{\partial C(t,S_{t^+})}{\partial t}\delta t+\frac{\partial C(t,S_{t^+})}{\partial S_{t^+}}\delta (S_{t^+})\nonumber\\
&+&\frac{1}{2}\frac{\partial^2C(t,S_{t^+})}{\partial S_{t^+}^2}(\delta S_{t^+})^2+O\big((\delta t)^{\frac{3}{2}}\sqrt{\log (\delta t)^{-1}}\big)\Big)\Big]-(1-\lambda\delta t)E\Big[\frac{kS_t}{2}|\delta \psi(t)|\Big]\nonumber\\
&-&\lambda\delta tE\Big[\frac{kS_t}{2}\Big|\psi(t,S_{t^+})-\psi(t,S_t)+\frac{\partial \psi(t,S_{t^+})}{\partial t}\delta t+\frac{\partial \psi(t,S_{t^+})}{\partial S_{t^+}}\delta (S_{t^+})\nonumber\\
&+&\frac{1}{2}\frac{\partial^2\psi(t,S_{t^+})}{\partial S_{t^+}^2}(\delta S_{t^+})^2+O\big((\delta t)^{\frac{3}{2}}\sqrt{\log (\delta t)^{-1}}\big)\Big|\Big]
=rP_t\delta t+O\big((\delta t)^2\big)\nonumber,
\label{eq:a18}
\end{eqnarray}
where the current stock price $S_t$ is given.
Since
\begin{eqnarray}
E\Big[\frac{kS_t}{2}|\delta \psi(t)|\Big]&=&\frac{(1-\lambda \delta t)kS_t}{2}E|\delta \psi(t)|+\frac{(\lambda\delta t)kS_t}{2}\Big|\psi(t,S_{t^+})\nonumber\\
&-&\psi(t,S_t)
+\frac{\partial \psi(t,S_{t^+})}{\partial t}\delta t+\frac{\partial \psi(t,S_{t^+})}{\partial S_{t^+}}\delta (S_{t^+})\nonumber\\
&+&\frac{1}{2}\frac{\partial^2\psi(t,S_{t^+})}{\partial S_{t^+}^2}(\delta S_{t^+})^2+O\big((\delta t)^{\frac{3}{2}}\sqrt{\log (\delta t)^{-1}}\big)\Big|\nonumber\\
&=&\frac{(1-\lambda \delta t)kS_t}{2}E\Big|\frac{\partial \psi(t,S_t)}{\partial t}\delta t+\frac{\partial \psi(t,S_t)}{\partial S_t}\delta S_t\nonumber\\
&+&\frac{1}{2}\frac{\partial^2\psi(t,S_t)}{\partial S_t^2}(\delta S_t)^2
+O\big((\delta t)^{\frac{3}{2}}\sqrt{\log (\delta t)^{-1}}\big)\Big|\nonumber\\
&+&\frac{(\lambda\delta t)kS_t}{2}\Big|\psi(t,S_{t^+})
-\psi(t,S_t)+\frac{\partial \psi(t,S_{t^+})}{\partial t}\delta t\nonumber\\
&+&\frac{\partial \psi(t,S_{t^+})}{\partial S_{t^+}}\delta (S_{t^+})+\frac{1}{2}\frac{\partial^2\psi(t,S_{t^+})}{\partial S_{t^+}^2}(\delta S_{t^+})^2\nonumber\\
&+&O\big((\delta t)^{\frac{3}{2}}\sqrt{\log (\delta t)^{-1}}\big)\Big|\nonumber\\
&\approx&\frac{kS_t^2}{2}\Big|\frac{\partial \psi}{\partial S_{t^+}}\Big|E|\sigma\delta B(t)+\sigma_H\delta B_H(t)|\nonumber\\
&+&\frac{k\lambda S_t\delta t}{2}|\psi(t,S_{t^+})-\psi(t,S_t)|+O(\delta t)\nonumber\\
&=&\frac{kS_t^2}{2}\Big|\frac{\partial \psi}{\partial S_t}\Big|\sqrt{\frac{2}{\pi}\Big(\sigma^2\delta t+\sigma_H^2(\delta t)^{2H}\Big)}\nonumber\\
&+&\frac{k\lambda S_t\delta t}{2}|\psi(t,S_{t^+})-\psi(t,S_t)|+O(\delta t),
\label{eq:a19}
\end{eqnarray}
and $\psi=\frac{\partial C}{\partial S_t}$, from equations (\ref{eq:a1})-(\ref{eq:a18}), we can get

\begin{eqnarray}
&&\Big[\frac{\partial C}{\partial t}+rS_t\frac{\partial C}{\partial S_t}+\frac{S_t^2}{2}(\sigma^2+\sigma_H^2(\delta t)^{2H-1})\frac{\partial^2 C}{\partial S_t^2}\nonumber\\
&&-rC+\lambda E[C(t,J_tS_t)-C(t,S_t)]-\lambda E[J-1]S_t\frac{\partial C}{\partial S_t}\nonumber\\
&&+\frac{kS_t^2}{2}\sqrt{\frac{2}{\pi}\big(\frac{\sigma^2}{\delta t} +\sigma_H^2(\delta t)^{2H-2}\big)}\Big|\frac{\partial^2 C}{\partial S_t^2}\Big|\Big]\delta t+O(\delta t)=0.
\label{eq:a20}
\end{eqnarray}
Hence, we assume that
\begin{eqnarray}
&&\frac{\partial C}{\partial t}+rS_t\frac{\partial C}{\partial S_t}+\frac{S_t^2}{2}(\sigma^2+\sigma_H^2(\delta t)^{2H-1})\frac{\partial^2 C}{\partial S_t^2}\nonumber\\
&&-rC+\lambda E[C(t,J_tS_t)-C(t,S_t)]-\lambda E[J-1]S_t\frac{\partial C}{\partial S_t}\nonumber\\
&&+\frac{kS_t^2}{2}\sqrt{\frac{2}{\pi}\big(\frac{\sigma^2}{\delta t} +\sigma_H^2(\delta t)^{2H-2}\big)}\Big|\frac{\partial^2 C}{\partial S_t^2}\Big|=0.
\label{eq:a21}
\end{eqnarray}

Note that the term $\frac{kS_t^2}{2}\sqrt{\frac{2}{\pi}\big(\frac{\sigma^2}{\delta t} +\sigma_H^2(\delta t)^{2H-2}\big)}$ is nonlinear, except when $\Gamma=\frac{\partial^2 C}{\partial S_t^2}$ does not change sign for all $S_t$. Since it represents the degree of mishedging of the portfolio, it is not surprising to observe that $\Gamma$ is involved in the transaction cost term. We may rewrite equation (\ref{eq:a21}) in the form which resembles the Merton equation:
\begin{eqnarray}
&&\frac{\partial C}{\partial t}+rS_t\frac{\partial C}{\partial S_t}+\frac{S_t^2\widehat{\sigma}^2}{2}\frac{\partial^2 C}{\partial S_t^2}-rC\nonumber\\
&&+\lambda E[C(t,J_tS_t)-C(t,S_t)]-\lambda E[J-1]S_t\frac{\partial C}{\partial S_t}=0.
\label{eq:a22}
\end{eqnarray}

where $E[J-1]=e^{\mu_J+\frac{\sigma_J^2}{2}}-1$ and the implied volatility is given by
\begin{eqnarray}
\widehat{\sigma}^2=\sigma^2+\sigma_H^2(\delta t)^{2H-1}+k\sqrt{\frac{2}{\pi}\left(\frac{\sigma^2}{\delta t}+\sigma_H^2(\delta t)^{2H-2}\right)}sign(\Gamma).
\label{eq:a23}
\end{eqnarray}

If $\widehat{\sigma}^2$, equation (\ref{eq:a22}) becomes mathematically ill-posed. This occurs when $\Gamma<0$ and $\delta t\rightarrow 0$. However,
it is known that $\Gamma$ is always positive for the simple European call and put options in the absence of transaction costs. If
we postulate the same sign behaviour for $\Gamma$ in the presence of transaction costs, equation (\ref{eq:a22}) becomes linear under such an
assumption so that the Merton formula becomes applicable except that the modified volatility $\widehat{\sigma}$ should be used as the
volatility parameter. Moreover, if $\Gamma>0$ from equation (\ref{eq:a22}) we obtain

\begin{eqnarray}
C(t,S_t)=\sum_{n=0}^\infty\frac{e^{-\lambda'(T-t)}(\lambda'(T-t))^n}{n!}\Big[S_t\phi(d_1)-Ke^{-r(T-t)}\phi(d_2)\Big]\nonumber,
\label{eq:a24}
\end{eqnarray}
where
\begin{eqnarray}
d_1=\frac{\ln\left(\frac{S_t}{K}\right)+r_n(T-t)+\frac{\sigma_n^2}{2}(T-t)}{\sigma_n\sqrt{T-t}},\quad d_2=d_1-\sigma_n\sqrt{T-t}\nonumber,
\label{eq:a25}
\end{eqnarray}
\begin{eqnarray}
\lambda'=\lambda E(J)=\lambda e^{\mu_J+\frac{\sigma_J^2}{2}},\quad{\sigma_n}^2=\widehat{\sigma}^2+\frac{n\sigma_J^2}{T-t}\nonumber,
\label{eq:a26}
\end{eqnarray}
\begin{eqnarray}
r_n=r-\lambda E(J-1)+\frac{n\ln E(J)}{T-t}=r+\lambda (e^{\mu_J+\frac{\sigma_J^2}{2}}-1)+\frac{n(\mu_J+\frac{\sigma_J^2}{2})}{T-t}\nonumber,
\label{eq:a27}
\end{eqnarray}
\begin{eqnarray}
\widehat{\sigma}^2=\sigma^2+\sigma_H^2(\delta t)^{2H-1}+k\sqrt{\frac{2}{\pi}\Big(\frac{\sigma^2}{\delta t}+\sigma_H^2(\delta t)^{2H-2}\Big)}sign(\Gamma)\nonumber,
\label{eq:a28}
\end{eqnarray}
sign$(\Gamma)$ is the signum function of $\frac{\partial^2C}{\partial S_t^2}$, $\delta t$ is a small and fixed time-step, $k$ is the transaction costs and $\phi(.)$ is the cumulative normal distribution.

\bibliographystyle{siam}
\bibliography{../../reference}

\end{document}